\documentclass[prd,11pt,nofootinbib,preprint,showpacs,showkeys]{revtex4-1}
\usepackage[mathcal]{euscript}
\usepackage{latexsym}
\usepackage{amsmath}
\usepackage{graphicx}
\usepackage{hyperref}   
\usepackage{verbatim}   
\usepackage{subfigure}  
\usepackage{bm}
\usepackage{graphicx}
\usepackage{amssymb}
\usepackage{bbm}
\usepackage{slashed}
\usepackage{amsfonts}
\usepackage{euscript}
\usepackage{mathrsfs} 
\usepackage{bbding}
\usepackage{pifont}
\usepackage{cancel}

\usepackage[utf8]{inputenc}
\usepackage[usenames,dvipsnames]{color}

\makeindex

\begin{document}

\title{Exponential fading to white of black holes in quantum gravity}

\author{Carlos Barcel\'o}
\email{carlos@iaa.es}
\affiliation{Instituto de Astrof\'{\i}sica de Andaluc\'{\i}a (IAA-CSIC), Glorieta de la Astronom\'{\i}a, 18008 Granada, Spain}
\author{Ra\'ul Carballo-Rubio}
\email{raulc@iaa.es / Now at the Department of Mathematics \& Applied Mathematics, University of Cape Town.}
\affiliation{Instituto de Astrof\'{\i}sica de Andaluc\'{\i}a (IAA-CSIC), Glorieta de la Astronom\'{\i}a, 18008 Granada, Spain}
\author{Luis J. Garay}
\email{luisj.garay@ucm.es}
\affiliation{Departamento de F\'{\i}sica Te\'orica II, Universidad Complutense de Madrid, 28040 Madrid, Spain}
\affiliation{Instituto de Estructura de la Materia (IEM-CSIC), Serrano 121, 28006 Madrid, Spain}

\begin{abstract}{Quantization of the gravitational field may allow the existence of a decay channel of black holes into white holes with an explicit time-reversal symmetry. The definition of a meaningful decay probability for this channel is studied in spherically symmetric situations. As a first nontrivial calculation, we present the functional integration over a set of geometries using a single-variable function to interpolate between black-hole and white-hole geometries in a bounded region of spacetime. This computation gives a finite result which depends only on the Schwarzschild mass and a parameter measuring the width of the interpolating region. The associated probability distribution displays an exponential decay law on the latter parameter, with a mean lifetime inversely proportional to the Schwarzschild mass. In physical terms this would imply that matter collapsing to a black hole from a finite radius bounces back elastically and instantaneously, with negligible time delay as measured by external observers. These results invite to reconsider the ultimate nature of astrophysical black holes, providing a possible mechanism for the formation of black stars instead of proper general relativistic black holes. The existence of both this decay channel and black stars can be tested in future observations of gravitational waves.
}
\end{abstract}
\maketitle
\flushbottom

\section{Introduction}

The construction of a well-defined and physically meaningful theory of quantum gravity is the aim of different research programs \cite{Carlip2001}. Any attempt to describe the quantum properties of the gravitational field must  inevitably make a series of leaps, by means of the consideration of additional mathematical structures or self-consistent rules that go further than the classical description of spacetime geometry embodied in general relativity. The presumed extreme weakness of observable effects associated with quantization makes this research field highly speculative from an empirical standpoint. Whatever approach (or just partial construction) showing hints that quantum gravity effects may be larger, and hence testable in the near future, would be greatly valuable in that it might help to break this vicious circle.

A well-known consequence of partial (i.e., semiclassical) quantization is the destruction of the stability of black holes through an extremely faint, but nonzero, evaporation due to the emission of Hawking radiation \cite{Hawking1974,Unruh1976}. In a series of papers \cite{Barcelo2014,Barcelo2015,Barcelo2016u,Barcelo2016} the authors have proposed that black holes might decay instead through a time-symmetric decay channel that outdrives Hawking evaporation as the dominant channel once non-perturbative quantum gravity effects are taken into account, leading to white holes as the end product.\footnote{See \cite{Gambini2014,Corichi2015,Bambi2016,DeLorenzo2015} for other qualitatively different scenarios that include the formation of white holes.} In this picture, black holes formed in the gravitational collapse of massive stars from a finite radius would just represent short-lived configurations that are followed almost instantaneously by the bounce of the collapsing star back to its initial radius, at least in an idealized situation with no dissipation. Including dissipation would lead to dampened oscillations and eventual stabilization of the bouncing structure in a black ultra-compact horizonless object. This dynamical process points towards the formation of black stars, supported by quantum vacuum effects \cite{Barcelo2007,Visser2009a}. As explained in the conclusions, the corresponding gravitational wave signature associated with the gravitational collapse of massive stars would be substantially different from the templates dictated by general relativity. Also black stars present characteristics that may permit to distinguish them from black holes \cite{Barcelo2015}, in particular in binary coalescences \cite{Cardoso2016b}. The collection of additional gravitational wave data, following the first events observed \cite{Abbott2016,Abbott2016b}, will permit to put to test these features.

Previous work by the authors has focused on the properties of spacetime geometries describing in effective terms the decay of black holes into white holes in a given interval of time. While the knowledge of these effective geometries represents a natural starting point to study the physical implications of this picture, such as the specific form of the associated gravitational wave signature, it remains to be seen that this decay channel becomes indeed allowed in quantum gravity. Given the mathematical subtleties involved and the very vagueness on the precise meaning of the latter concept, this question is hard to answer in an exhaustive way. The present paper communicates two nontrivial results related to this question: (1) It is possible to use these effective geometries to obtain a measure of the probability that the time-symmetric decay of black holes into white holes takes place in an arbitrary time interval, and (2) The probability distribution on this time interval takes the form of an exponential decay law, with a mean lifetime of black holes which is in complete accordance with previous considerations by the authors.

The contents of this paper are up to certain extent inspired by the Euclidean approach to quantum gravity \cite{Gibbons1976,Gibbons1993}. This approach aims to construct a consistent path integral through the consideration of the analytic continuation of Lorentzian metrics to Riemannian metrics, hence avoiding oscillatory integrals that appear in the Lorentzian case. This motivation leads to the following formal expression for the amplitude between two configurations $h_-$ and $h_+$ of the spacetime geometry at hypersurfaces $\Sigma_-$ and $\Sigma_+$, respectively:
\begin{equation}
\langle h_+,\Sigma_+|h_-,\Sigma_-\rangle=\frac{1}{\mathscr{N}}\int_{g(\Sigma_-)=h_-}^{g(\Sigma_+)=h_+}\mathscr{D}g\exp(-\mathscr{S}_{\rm EH}[g]).\label{eq:famp}
\end{equation}
Here $\mathscr{S}_{\rm EH}[g]$ is the Einstein-Hilbert action of a Euclidean geometry $g$ satisfying the boundary conditions specified at $\Sigma_-$ and $\Sigma_+$, $\mathscr{D}g$ is the measure on the configuration space of Euclidean geometries, and $\mathscr{N}$ is a normalization constant. Dealing with Euclidean geometries helps to improve the convergence properties of the path integral, though Eq. \eqref{eq:famp} is still formal in that it is not well-defined in the absence of further considerations \cite{Gibbons1993}.

In this paper we define a suitable version of Eq. \eqref{eq:famp} that is valid for a symmetry-reduced situation. This functional integral will be defined as the sum over a given set of geometries, invariant under time-reversal and spatial rotations, and interpolating between a black-hole geometry at $\Sigma_-$ and a white-hole geometry at $\Sigma_+$. In this integration, we include all the geometries interpolating between black-hole geometries and white-hole geometries that use a single-variable interpolating function. The results obtained show that these geometries provide the dominant contribution in the relevant region of the parameter space, hence proving that this approximation is self-consistent. On the other hand, the hypersurfaces $\Sigma_\pm$ are arbitrary (but for a set of minimal technical assumptions), which shows explicitly the generality of the results obtained. These considerations permit to go from the ambiguous quantity in Eq. \eqref{eq:famp} to a tractable expression which is evaluated in similar terms to one-dimensional quantum-mechanical problems, and indeed displays a finite value.

After the evaluation of the relevant functional integrals, the results that follow are analyzed. The resulting probability distribution exhibits an exponential decay on one of the parameters measuring the size of the interpolating region $\Gamma$. We show that the mean lifetime in this exponential-decay law is inversely proportional to the Schwarzschild mass, which in terms of external observers means that the decay of black hole into white holes is extremely fast (indeed the fastest possible in practical terms). This result, together with the use of Euclidean path integrals and the lack of a classical solution joining the black-hole and white-hole geometries at $\Sigma_-$ and $\Sigma_+$, may suggest the interpretation of this phenomenon as a tunneling effect. Following this interpretation the effective geometries in the interpolating region $\Gamma$, not being vacuum solutions of the Einstein field equations, would correspond to the classically forbidden configurations.

\section{Interpolating geometries}

Our aim is to evaluate a functional integral over a set of (Euclidean) interpolating geometries that describe a smooth transition between black-hole and white-hole geometries. The first step in this program is naturally the definition of the Euclidean geometries to be considered. Natural units $G=\hbar=c=1$ are used in the following.

Let us start recalling the form of the time-symmetric Lorentzian geometries interpolating in a bounded region of spacetime $\Gamma$ between a black hole with mass $M$ and its time-reversal solution corresponding to a white hole. The existence of these geometries was discussed in \cite{Barcelo2014,Barcelo2015}, where it was shown that the interpolating region has to extend further than the Schwarzschild radius $2M$. The fine details concerning the explicit construction of these geometries can be read in \cite{Barcelo2016u}. Here we just mention briefly the main results of these papers that are of direct relevance for the present discussion.

Before entering in specific details, we want to stress that we are assuming that there is no microscopic arrow of time in quantum gravity. This motivates studying a time-symmetric decay channel. This is shared by many approaches to the construction of a theory to quantum gravity, but for certain notable exceptions \cite{Wald1980,Penrose1980,Liu1993,Barbour2014}. Most importantly, the lack of a microscopic arrow of time does not enter in contradiction with the T-violation observed in particle physics, as explained for instance in \cite{Quinn2009}.

The line element of the geometries to be considered can be written as
\begin{align}
g_{ab}\text{d}x^a\text{d}x^b=-\text{d}t^2+\frac{[\text{d}r-f(u)v(r)\text{d}t]^2}{1-2M/r_{\rm i}}+r^2\text{d}\Omega_2^{2}.\label{eq:metrics1}
\end{align}
Here we are using coordinates $x^a=(t,r,\theta,\varphi)$, $\text{d}\Omega_2^{2}$ is the line element of the $2$-sphere with angular coordinates $(\theta,\varphi)$, $r\leq r_{\rm i}$ is the radial coordinate, $v(r)=(2M/r_{\rm i}-2M/r)^{1/2}$ and
\begin{equation}
u=\frac{t}{\Delta (r)}.\label{eq:fform}
\end{equation}
As discussed below, $\Delta (r)$ is a fixed, but unspecified, function that describes the shape of the spacetime region $\Gamma$ interpolating between these two classical geometries. For $f(u)=\mp1$ one would recover two specific patches of the Schwarzschild solution going from $r=0$ to $r=r_{\rm i}$; these correspond, respectively, to the black-hole and white-hole patches in Painlev{\'e}-Gullstrand coordinates \cite{Martel2000}. The function $f(u)$ interpolates between these two limiting values so that Eq. \eqref{eq:metrics1} represents the geometry outside a stellar structure undergoing gravitational collapse and a subsequent time-symmetric bounce at $t=0$, though we will not need to consider explicitly the geometry inside the star. Due to this interpolation, the coordinates being used are not the usual ingoing or outgoing Painlev{\'e}-Gullstrand coordinates, but rather a non-trivial combination of them. The parameter $r_{\rm i}>2M$ that marks the extension of the interpolating region is coincident with the initial radius of the star (this identification between two in principle different quantities follows from the explicit construction of the interpolating geometries).

The only restriction on these interpolations is that they are effectively one-dimensional, which fixes their dependence on the variable $u$. The single-variable function $f(u)$ interpolates then between the values corresponding to the black-hole patch in the ingoing Painlev{\'e}-Gullstrand coordinates $[f(u)=-1$] and the white-hole patch in the outgoing Painlev{\'e}-Gullstrand coordinates [$f(u)=1$]. In order to make explicit the role played by the function $\Delta(r)$, let us define a subset $\Gamma$ of the $(t,r)$ coordinates as 
\begin{equation}
\Gamma=\{(t,r)|\,t\in[-\Delta(r),\Delta(r)],r\in[r_0,r_{\rm i}]\}.
\end{equation}
Here $r_{\rm 0}\sim M^{1/3}\ll r_{\rm i}$ is the radius in which the deviations from the classical collapsing solution due to quantization are of order unity (equivalently, the radius in which the spacetime curvature becomes Planckian), leading generally to an effective repulsive potential that modifies the spacetime geometry \cite{Frolov1979,Frolov1981,Hajicek2002,Ambrus2005,Hayward2005,Frolov2014,Frolov2016,Kawai2013,Bambi2013,Torres2015,Arfaei2016}. This radius is around $10^{13}$ in natural units for solar-mass stars, or $10^{-22}$ $\rm{m}$ in SI units. The boundaries of $\Gamma$ over which boundary conditions on $f(u)$ are imposed are given by 
\begin{equation}
\Sigma_\pm=\{(t,r)|\,t=\pm\Delta (r),r\in[r_0,r_{\rm i}]\}.\label{eq:sigmadef}
\end{equation}
The boundary conditions to be satisfied by $f(u)$ take then the simple form $f(u=\pm1)=\pm1,\qquad f'(u=\pm1)=0$. Both conditions guarantee that the interpolating region matches smoothly the black-hole and white-hole geometries at $\Sigma_-$ and $\Sigma_+$, respectively. Note that time-reversal invariance implies $f(-u)=-f(u)$, hence $f(0)=0$. Due to time-reversal symmetry, each geometry in Eq. \eqref{eq:metrics1} is alternatively specified by a choice of function $f(u)$ for $u\in [0,1]$, such that $|f(u)|\leq1$ and satisfying the boundary conditions $f(0)=0$, $f(1)=1$ and $f'(1)=0$. These conditions define the functional space $\mathscr{F}$ on the interval $u\in[0,1]$.

The particular form of the function $\Delta (r)$ defines the hypersurfaces $\Sigma_\pm$ in Eq. \eqref{eq:sigmadef} or, in other words, the shape of the interpolating region $\Gamma$, and is therefore fixed. The degeneracy in the choice of $\Delta(r)$ cannot be reduced without further knowledge of quantum gravity. Hence we consider a general function $\Delta(r)$ from the beginning, and show explicitly that the results below are robust against changes in this function. The only technical conditions imposed on $\Delta(r)$ is that it is a concave downwards function that decreases monotonically outwards until reaching a zero value at $r=r_{\rm i}$, namely $\Delta(r_{\rm i})=0$. Under these conditions, the (maximum) value of this function $\Delta_0=\Delta(r_0)$ controls the size of $\Gamma$ and hence the duration of the transition between the black-hole and white-hole geometries. It is useful to define a dimensionless variable $x=r/r_{\rm i}$ and a dimensionless function $I(x)=\Delta(x)/\Delta_0$.

We now define a rule to associate a unique Euclidean metric $\bar{g}_{ab}$ to each Lorentzian metric $g_{ab}$ in Eq. \eqref{eq:metrics1}, and hence a weight in the Euclidean path integral \eqref{eq:famp}. Given the structure of the spacetime line element we are considering, it is enough to follow the usual prescription of analytical continuation to imaginary values of the temporal coordinate $t$ (see, e.g., \cite{Mukhanov2007}). Let us consider a new temporal coordinate $T=it$ where $i$ is the imaginary unit; for real $t$ the coordinate $T$ takes purely imaginary values. The standard prescription we follow then is replacing $t$ by $T$ and performing the analytic continuation of the functions of the time $t$ so that these functions take real values for $T$ real. Specifically, we define $if(-it/\Delta)=-\bar{f}(T/\Delta)$, where now $\bar{f}(T/\Delta)$ is taken as a real function of the variable $\bar{u}=T/\Delta(r)$. This leads to the following Euclidean metric $\bar{g}_{ab}$ for each Lorentzian metric $g_{ab}$ in Eq. \eqref{eq:metrics1}:
\begin{align}
\bar{g}_{ab}\text{d}x^a\text{d}x^b=\text{d}T^2+\frac{[\text{d}r-\bar{f}(\bar{u})v(r)\text{d}T]^2}{1-2M/r_{\rm i}}+r^2\text{d}\Omega_2^{2}.\label{eq:metricsw2}
\end{align}
In the following we drop the overline on the metric $\bar{g}_{ab}$, the function $\bar{f}(\bar{u})$ and its variable $\bar{u}$, keeping in mind that we will be working in Euclidean signature with line element \eqref{eq:metricsw2}.

\section{Probability amplitude}

In this section we define and evaluate the probability amplitude for the transition between black-hole and white-hole geometries. The geometry in the interpolating region $\Gamma$, not being a vacuum solution of the Einstein field equations, leads to a nonzero value of the Einstein-Hilbert action $\mathscr{S}_{\rm EH}[g]=\int_\Gamma\bm{\epsilon}\,R(g)/16\pi$, where $\bm{\epsilon}$ is the Riemannian volume form and $R(g)$ the Ricci scalar of the Euclidean metric in Eq. \eqref{eq:metricsw2}. This nonzero value, which can be evaluated directly from the form of the metric in Eq. \eqref{eq:metricsw2}, permits to define a nontrivial weight $\exp(-\mathscr{S}_{\rm EH}[g])$ which is understood as the probability measure for a given interpolating geometry. This leads to the natural definition of the probability amplitude between black-hole and white-hole geometries at $\Sigma_-$ and $\Sigma_+$ as the functional integral of this exponential factor over all the possible interpolating geometries satisfying the boundary conditions. The Einstein-Hilbert action evaluated on the interpolating geometries can be recast in a remarkably simple form after lengthy manipulations (detailed in \ref{sec:app}), leading to the remarkably simple equation
\begin{align}
\langle{\rm WH|BH}\rangle_{M,\Delta_0}&=\frac{1}{\mathscr{N}'}\int_{\mathscr{F}}\mathscr{D}f\,\exp(-\mathscr{S}_{\rm EH}[g])\nonumber\\
&=\frac{1}{\mathscr{N}}\exp\left(-\frac{k_IM\Delta_0}{\sqrt{1-2M/r_{\rm i}}}\right)\int_{\mathscr{F}}\mathscr{D}f\,\exp\left[-\frac{M\Delta_0}{\sqrt{1-2M/r_{\rm i}}}\int_0^1\text{d}u\,f^2(u)\right].\label{eq:tamp}
\end{align}
This amplitude is defined in terms of a Gaussian functional integral over single-variable functions $f(u)\in\mathscr{F}$. The numerical constant $k_I$ is a function of the area spanned by $I(x)=\Delta(x)/\Delta_0$, taking values in the interval $k_I\in[1,3]$. In Eq. \eqref{eq:tamp} as well as in the following equations, only the leading order in the extremely small parameter $r_0/r_{\rm i}$ is written explicitly. Subdominant terms on the parameter $r_0/r_{\rm i}$ in the Einstein-Hilbert action have been explicitly considered and shown to be irrelevant for our conclusions. Also note that, in Eq. \eqref{eq:tamp}, the pieces of the classical action that do not display dependence on $\Delta_0$ have been absorbed in the definition of the normalization constant.

The evaluation of the Gaussian functional integral in Eq. \eqref{eq:tamp} can be accomplished considering a discretization of the interval $[0,1]$ defined in terms of a set of points $\{u_i\}_{i=1}^{N+2}$ and taking the limit $N\rightarrow\infty$ after performing the necessary calculations. If we define $f_i=f(u_i)\in[0,1]$, $i=1,...,N+2$, the integral inside the exponential in Eq. \eqref{eq:tamp} is discretized as
\begin{equation}
\int_0^1\text{d}u\,f^2(u)\longrightarrow\frac{1}{N+2}\sum_{i=1}^{N+2}f_i^2=\frac{2}{N+2}+\frac{1}{N+2}\sum_{i=1}^{N}f_i^2,
\end{equation}
where we have used $f_{N+1}=f_{N+2}=1$ as a consequence of the boundary conditions satisfied by $f(u)$. The $N$-dimensional measure for the discretized functional integral is defined as $\prod_{i=1}^{N}\text{d}f_i$. The Gaussian functional integral in Eq. \eqref{eq:tamp} arises taking the $N\rightarrow\infty$ limit in this discretization. This choice of measure is the natural one given the parametrization of the interpolating geometries. Also this definition is consistent with the potential existence of diffeomorphisms preserving the structure of the line element \eqref{eq:metrics1} and reduce to the identity transformation at the boundaries $\Sigma_{\pm}$. Under these ``residual'' diffeomorphisms, the function $f(u)$ may be non-trivially transformed into another function $f'(u')$ where $u'$ is defined in the same way as $u$. The corresponding path integral measure would be given by $\prod_{i=1}^{N}\text{d}f'_i$, and the overall transition amplitude in Eq. \eqref{eq:tamp} is indeed invariant.

The discretized version of the functional integral in Eq. \eqref{eq:tamp} is then expressed in terms of the product of $N$ integrals of the form
\begin{equation}
\int_0^1\text{d}f_i\exp\left(-\frac{\lambda f_i^2}{N+2}\right)=\frac{\sqrt{\pi}}{2}\sqrt{\frac{N+2}{\lambda}}\mbox{erf}\left(\sqrt{\frac{\lambda}{N+2}}\right),
\end{equation}
where $\mbox{erf}(x)$ is the error function with the usual normalization \cite{Abramowitz1964} and $\lambda=M\Delta_0/\sqrt{1-2M/r_{\rm i}}$. Using the Taylor expansion $\mbox{erf}(x)=2[x-x^3/3+\mathscr{O}(x^5)]/\sqrt{\pi}$ and the definition of the exponential as the limit $\exp(x)=\lim_{N\rightarrow\infty}[1+x/N+\mathscr{O}(1/N^2)]^N$ permits to show that
\begin{align}
\int_{\mathscr{F}}\mathscr{D}f\,\exp\left[-\lambda\int_0^1\text{d}u\,f^2(u)\right]=\exp\left(-\lambda/3\right).\label{eq:gint12}
\end{align}
Hence the probability amplitude \eqref{eq:tamp} takes the finite value
\begin{align}
\langle {\rm WH|BH}\rangle_{M,\Delta_0}=\frac{1}{\mathscr{N}}\exp\left(-\frac{(k_I+1/3)M\Delta_0}{\sqrt{1-2M/r_{\rm i}}}\right).\label{eq:tampf1}
\end{align}
It is important to remark that the transition amplitude above has been evaluated exactly. There is no need to use the well-known saddle-point method or any other approximation method. Of course, this does not affect the interpretation of this transition amplitude as giving the probability of the decay of a black hole into a white hole, which stems from its very definition (and not the specific approach chosen for its evaluation).
 
\section{Exponential decay}

Eq. \eqref{eq:tampf1} takes the form that would be expected from a tunneling amplitude, with $\Delta_0$ measuring the width of the classically forbidden region. The square of Eq. \eqref{eq:tampf1} gives then the probability density of tunneling between black-hole and a white-hole geometries for a specific value of $\Delta_0\in[0,\infty)$. To obtain the probability that this transition takes place in a finite interval $[0,\Delta_0]$ it is then necessary to evaluate the one-dimensional integral
\begin{equation}
P_{\rm BH\,\triangleright WH}(M,\Delta_0)=\int_0^{\Delta_0}\text{d}\Delta_0'\,|\langle {\rm WH|BH}\rangle_{M,\Delta_0'}|^2,\label{eq:finprob0}
\end{equation}
with the normalization constant $\mathscr{N}$ fixed by the normalization condition $P_{\rm BH\,\triangleright WH}(M,\infty)=1$. The evaluation of the integral \eqref{eq:finprob0} from Eq. \eqref{eq:tampf1} is straightforward, leading to the following exponential decay law:
\begin{equation}
P_{\rm BH\,\triangleright WH}(M,\Delta_0)=1-\exp\left(-\frac{2(k_I+1/3)M\Delta_0}{\sqrt{1-2M/r_{\rm i}}}\right).\label{eq:finprob}
\end{equation}
The standard normalization of Eq. \eqref{eq:finprob} ensures, from a mathematical standpoint, that the transition will always take place if one waits long enough (i.e., infinite time). This does not imply though that the decay should take place physically: if it happens to have a very long characteristic time scale, it would be disrupted by other effects; for instance, by the evaporation due to the emission of Hawking radiation (which is not taken into account in the evaluation of the probability amplitude above due to the assumption of time-reversal invariance), or more drastically by the existence of white hole instabilities as discussed in \cite{Barcelo2016}. Given the exponential decay form of Eq. \eqref{eq:finprob}, the quantity that determines whether this transition is physical is the mean lifetime $\tau$ defined as usual. Taking into account that $k_I+1/3$ is always positive and of order unity, it will be generally

\begin{equation}
\tau\simeq\frac{\sqrt{1-2M/r_{\rm i}}}{2M}\leq\frac{1}{2M}.\label{eq:mean}
\end{equation}

\section{Characteristic time scale}

The parameter $\Delta_0$ is tightly connected to a natural definition of the lifetime of the black hole, namely the time $\mathscr{T}_{r_{\rm i}}$ that the bounce of the star takes as measured by an external observer situated at the initial radius of the collapsing star $r=r_{\rm i}$ \cite{Barcelo2014,Barcelo2015,Barcelo2016u}, or equivalently the same time interval as measured by asymptotic observers at $r\rightarrow\infty$, $\mathscr{T}_\infty$. The coordinate invariant quantity $\mathscr{T}_{r_{\rm i}}$ is given by twice (due to-time reversal symmetry) the classical collapsing proper time from $r_{\rm i}$ to $r_0$, $\mathscr{T}^0_{r_{\rm i}}$ (obtained in a first approximation using the Oppenheimer-Snyder model \cite{Oppenheimer1939}), plus an additional term proportional to $\Delta_0$. This can be evaluated explicitly from the line element \eqref{eq:metrics1}:
\begin{equation}
\mathscr{T}_{r_{\rm i}}=\mathscr{T}^0_{r_{\rm i}}+2\Delta_0.\label{eq:trmi}
\end{equation}
The latter term, which could be considered as the quantum-mechanical part, increases linearly with the value of $\Delta_0\in[0,\infty)$, being zero when $\Delta_0=0$. The value $\Delta_0=\tau$ given in Eq. \eqref{eq:mean} implies that this contribution is negligible  for macroscopic stars, so that the bounce is an elastic process which does not display an appreciable time delay to be added to $\mathscr{T}^0_{r_{\rm i}}$.

The same conclusion is valid for asymptotic observers at spatial infinity which measure the time interval $\mathscr{T}_{\infty}$, straightforwardly obtained from Eq. \eqref{eq:trmi} using the multiplicative redshift factor characteristic of the Schwarzschild solution $1/\sqrt{1-2M/r_{\rm i}}$. Eq. \eqref{eq:mean} implies then that the contribution to be added to the classical bouncing time scales as $1/2M$ for asymptotic observers. In simplified terms, Eq. \eqref{eq:mean} should be read as pointing to black holes as extremely unstable objects. Let us stress that these results cannot be extrapolated to horizonless objects, due to the use of boundary geometries at $\Sigma_{\mp}$ that correspond to vacuum solutions of the Einstein field equations in the entire range $r\in[r_0,r_{\rm i}]$. In astrophysical scenarios, $r_{\rm i}$ is given by a $\mathscr{O}(1)$ multiple of the Schwarzschild radius $2M$, corresponding typically to the radius of a (proto-)neutron star \cite{Woosley2006}. This leads to a black hole lifetime $\mathscr{T}_{\infty}$ roughly proportional to $M+\mathscr{O}(1/M)$, which is to be compared with the extremely larger evaporation time through Hawking evaporation, proportional to $M^3$. 

\section{Conclusions}

Following these results, the time-symmetric decay into white holes would outdrive Hawking evaporation, becoming the dominant decay channel for black holes in quantum gravity. Unless there exists some (symmetry) principle forbidding this decay channel, it will dominate the physics of black holes. It is interesting to remark that classical and semiclassical arguments forbid longer time scales, but allow this time-symmetric decay as long as its characteristic time scale is linear in $M$  \cite{Barcelo2016,DeLorenzo2015}. We find the convergence of arguments of diverse nature concerning the time scale of the time-symmetric decay remarkable, and a strong incentive to further explore this phenomenon.

The main contribution of this paper is the evaluation of a well-defined probability amplitude for the time-symmetric decay of black holes into white holes, and the subsequent characteristic time scale. Evaluation of transition amplitudes in quantum gravity is notably difficult, so that we have considered a natural reduction of the number of degrees of freedom involved, through the use of single-variable interpolating functions in the line element \eqref{eq:metricsw2}. That this reduction permits to carry out an exact analytical treatment is noteworthy, as also are the results obtained, namely an exact exponential decay law and a mean lifetime fixed by the dynamics of general relativity to be given by Eq. \eqref{eq:mean}. It should be possible to check whether these features survive in different approaches and quantization schemes \cite{Campiglia2016,Olmedo2016,BenAchour2016,Christodoulou2016}; let us stress that time scales linear in $M$ have been indeed shown to arise in the canonical quantization of null shells \cite{Ambrus2005}.

In physical terms, our results point that quantum gravity effects in black holes may be more dramatic than expected, opening an observational window that is yet to be explored. In the approximation considered here which neglects dissipative effects (which is indeed a reasonable approximation due to the short time scale of the process), the collapse of a massive star from an initial radius $r_{\rm i}$ would be followed by an infinite number of cycles in which the star bounces back to $r_{\rm i}$. Undoubtedly, dissipation has to be taken into account in order to elaborate a satisfactory physical picture, including for instance the very emission of gravitational waves. Dissipative effects will gradually shrink the initial value of $r_{\rm i}$ down to $2M$. If this shrinking is slow enough (as it should be if dissipative effects represent a small correction), after a transient composed of a series of dampened oscillations, a stable configuration in the form of a black ultra-compact horizonless object with radius close to its gravitational radius can be stabilized due to quantum vacuum effects \cite{Barcelo2007}, and a black star is formed \cite{Visser2009a,Barcelo2010}. Hence the decay of black holes into white holes opens the possibility for the formation of black stars instead of black holes. Both electromagnetic and gravitational wave observations leave wiggle room for black ultra-compact horizonless objects of this kind representing the ultimate nature of astrophysical black holes, as discussed respectively in \cite{Abramowicz2002,Abramowicz2016} and  \cite{Cardoso2016} (see also the general discussion in \cite{Visser2009}). Only time (and future observations) will tell whether or not this possibility stands further theoretical and experimental scrutiny.

In this regard, gravitational wave observations will play an important role. Typical values of $r_{\rm i}$ makes detection of this decay channel by means of electromagnetic radiation difficult, due to the interference with many other physical processes. However, macroscopic quantum effects modifying the local geometry around the Schwarzschild radius would most probably leave a noticeable imprint (of transient nature) on gravitational radiation. Also the fact that the interpolating region connects two classical solutions (corresponding to the collapsing and expanding star, respectively) implies that there should be new emissions of gravitational radiation that are not present in classical general relativity, associated with the expansion of the (inhomogeneous) star across the white-hole geometry (inhomogeneities in the collapsing star will be generally amplified in the bounce as shown in cosmological scenarios \cite{Brizuela2009}). These features, together with the existence of a series of dampened oscillations characterized by the formation of short-lived black hole horizons, point to a gravitational wave signature showing periodic features that decay in time, and clearly different to the classical pattern in which only one cycle (or main signal) would be present. The existence or not of these reverberations in gravitational wave signals from collapse events provides a crucial way to distinguish between the formation of black holes or black stars. Regarding the coalescence of already formed black stars, the existence of a surface outside the would-be horizon may lead (depending on its reflective properties) to similar reverberant signatures \cite{Barcelo2015,Cardoso2016b}.

%------------------------------------
%Acknowledgements
%------------------------------------
 
\subsection*{Acknowledgments}
Financial support was provided by the Spanish MINECO through the projects FIS2011-30145-C03-01, FIS2011-30145-C03-02, FIS2014-54800-C2-1, FIS2014-54800-C2-2 (with FEDER contribution), and by the Junta de Andaluc\'{\i}a through the project FQM219. R.C-R. acknowledges support, at different stages of the elaboration of this work, from the Math Institute of the University of Granada (IEMath-GR), the research project MINECO-FEDER MTM2013-47828-C2-1-P, and the Claude Leon Foundation.

\appendix

\section{Evaluation of the action functional on the interpolating geometries \label{sec:app}}

In this appendix we evaluate the Einstein-Hilbert action on the interpolating geometries, namely the functional
\begin{equation}
\mathscr{S}_{\rm EH}[g]=\frac{1}{16\pi}\int_{\Gamma}\bm{\epsilon}\,R(g),
\end{equation}
where $R(g)$ is the Ricci scalar of the metric $g_{ab}$ with line element
\begin{align}
g_{ab}\text{d}x^a\text{d}x^b&=\left[1+\frac{f^2(u)}{1-2M/r_{\rm i}}\left(\frac{2M}{r}-\frac{2M}{r_{\rm i}}\right)\right]\text{d}T^2-\frac{2f(u)}{1-2M/r_{\rm i}}\sqrt{\frac{2M}{r}-\frac{2M}{r_{\rm i}}}\text{d}T\text{d}r\nonumber\\
&+\frac{\text{d}r^2}{1-2M/r_{\rm i}}+r^2\text{d}\Omega_2^{\ 2}.\label{eq:metricsw2app}
\end{align}
The functional form of $R(g)$ in terms of $f(u)$ is given by
\begin{align}
R(g)&=\frac{4M}{r_{\rm i}r^2}\left.\Bigg[1+f^2(u)-r(r_{\rm i}-r)\left(\frac{\partial f(u)}{\partial r}\right)^2-2(r_{\rm i}-2r)f(u)\frac{\partial f(u)}{\partial r}\right.\nonumber\\
&\left.-r(r_{\rm i}-r)f(u)\frac{\partial^2f(u)}{\partial r^2}-\frac{\displaystyle(3r_{\rm i}-4r)\partial f(u)/\partial T+2r(r_{\rm i}-r)\partial^2f(u)/\partial T\partial r}{2\displaystyle\sqrt{2M(1/r-1/r_{\rm i})}}\right].\label{eq:ricci1}
\end{align}
Using the form of the variable $u$ inside the interpolating function, namely $u=T/\Delta(r)$, leads to the relations
\begin{align}
&\frac{\partial f(u)}{\partial r}=-u\frac{\text{d}f}{\text{d}u}\frac{\text{d}\ln\Delta}{\text{d}r}, &\frac{\partial^2 f(u)}{\partial r^2}=u^2\frac{\text{d}^2f}{\text{d}u^2}\left(\frac{\text{d}\ln\Delta}{\text{d}r}\right)^2+u\frac{\text{d}f}{\text{d}u}\left[\left(\frac{\text{d}\ln\Delta}{\text{d}r}\right)^2-\frac{\text{d}^2\ln\Delta}{\text{d}r^2}\right],\nonumber\\
&\frac{\partial f(u)}{\partial T}=\frac{1}{\Delta}\frac{\text{d}f(u)}{\text{d} u}, &\frac{\partial^2(u)}{\partial T\partial r}=\frac{\text{d}}{\text{d}r}\left(\frac{1}{\Delta}\right)\frac{\text{d} f(u)}{\text{d} u}+u\frac{\text{d}^2 f(u)}{\text{d} u^2}\frac{\text{d}}{\text{d}r}\left(\frac{1}{\Delta}\right).
\end{align}
Due to spherical symmetry, the integration of the action functional is reduced to a two-dimensional integral on the $(t,r)$ variables. It will be useful to perform the integration in terms of the $(u,r)$ variables instead. The following relation holds for any function $h=h(T/\Delta,r)$:
\begin{equation}
\int_{r_0}^{r_{\rm i}}\text{d}r\int_{-\Delta(r)}^{\Delta(r)}\text{d}T\,h(T/\Delta,r)=\int_{-1}^{+1}\text{d}u\int_{r_0}^{r_i}\text{d}r\,\Delta (r)h(u,r).\label{eq:usef1}
\end{equation}
Note also that
\begin{equation}
g=\mbox{det}(g_{ab})=\frac{1}{\displaystyle1-2M/r_{\rm i}}.
\end{equation}
Let us split the different contributions in order to arrange conveniently the necessary calculations. Taking into account the $4\pi$ multiplicative factor coming from the integration over the angular variables, one has
\begin{align}
\mathscr{S}_{\rm EH}=\frac{1}{4\sqrt{1-2M/r_{\rm i}}}\sum_{n=1}^7\mathcal{E}^{(n)},
\end{align}
where:
\begin{align}
\mathcal{E}^{(1)}&=\int_{r_0}^{r_i}\text{d}r\,r^2\int_{-\Delta (r)}^{\Delta (r)}\text{d}T\,\frac{4M}{r_{\rm i}r^2},\\
\mathcal{E}^{(2)}&=\int_{r_0}^{r_i}\text{d}r\,r^2\int_{-\Delta (r)}^{\Delta (r)}\text{d}T\,\frac{4M}{r_{\rm i}r^2}f^2(u),\\
\mathcal{E}^{(3)}&=-\int_{r_0}^{r_i}\text{d}r\,r^2\int_{-\Delta (r)}^{\Delta (r)}\text{d}T\,\frac{4M}{r_{\rm i}r}(r_{\rm i}-r)\left(\frac{\partial f(u)}{\partial r}\right)^2,\\
\mathcal{E}^{(4)}&=-\int_{r_0}^{r_i}\text{d}r\,r^2\int_{-\Delta (r)}^{\Delta (r)}\text{d}T\,\frac{8M}{r_{\rm i}r^2}(r_{\rm i}-2r)f(u)\frac{\partial f(u)}{\partial r},\\
\mathcal{E}^{(5)}&=-\int_{r_0}^{r_i}\text{d}r\,r^2\int_{-\Delta (r)}^{\Delta (r)}\text{d}T\,\frac{4M}{r_{\rm i}r}(r_{\rm i}-r)f(u)\frac{\partial^2f(u)}{\partial r^2},\\
\mathcal{E}^{(6)}&=-\int_{r_0}^{r_i}\text{d}r\,r^2\int_{-\Delta (r)}^{\Delta (r)}\text{d}T\,\frac{2M}{r_{\rm i}r^2}\frac{\displaystyle(3r_{\rm i}-4r)}{\displaystyle\sqrt{2M(1/r-1/r_{\rm i})}}\frac{\partial f(u)}{\partial T},\\
\mathcal{E}^{(7)}&=-\int_{r_0}^{r_i}\text{d}r\,r^2\int_{-\Delta (r)}^{\Delta (r)}\text{d}T\,\frac{4M}{r_{\rm i}r^2}\frac{r(r_{\rm i}-r)}{\displaystyle\sqrt{2M(1/r-1/r_{\rm i})}}\frac{\partial^2f(u)}{\partial T\partial r}.
\end{align}
It is now a matter of evaluating these integrals, using Eq. \eqref{eq:usef1}. In order to extract the dependence of these quantities on the dimensionful parameters of the problem, let us define $x=r/r_{\rm i}$ and $I(x)=\Delta(x)/\Delta_0$.

\noindent
The first contribution is given by
\begin{equation}
\mathcal{E}^{(1)}=4M \Delta_0\int_{-1}^{+1}\text{d}u\int_{r_0/r_{\rm i}}^{1}\text{d}x\,I(x)=8M \Delta_0\int_{r_0/r_{\rm i}}^{1}\text{d}x\,I(x).\label{eq:cale1c}
\end{equation}
This equation shows that $\mathcal{E}^{(1)}$ depends only on two dimensionful constants: the Schwarzschild mass $M$ and $\Delta_0$. The dependence is moreover linear in the product $M\Delta_0$. The particular value of the numerical coefficient multiplying this quantity depends on the integrals on the right-hand side of Eq. \eqref{eq:cale1c}. Hence changes on the function $I(x)$, or equivalently $\Delta(r)$ will lead to modifications of dimensionless coefficients, which are of order unity. This structure is shared by all the terms $\mathcal{E}^{(n)}$ of the classical action.

\noindent
The evaluation of the second contribution is similar to the previous one:
\begin{equation}
\mathcal{E}^{(2)}= 4M \Delta_0\int_{-1}^{+1}\text{d}u\,f^2(u)\int_{r_0/r_{\rm i}}^{1}\text{d}x\,I(x).
\end{equation}
Again, we can see explicitly that changes of $\Delta(r)$ lead to modifications in the dimensionless numerical factors in the equation above.
\noindent
The third contribution reads
\begin{align}
\mathcal{E}^{(3)}&=-\int_{-1}^{+1}\text{d}u\int_{r_0}^{r_i}\text{d}r\,\Delta (r)\frac{4M}{r_{\rm i}}r(r_{\rm i}-r)\left(\frac{\partial f(T/\Delta)}{\partial r}\right)^2\nonumber\\
&=-\frac{4M}{r_{\rm i}}\int_{-1}^{+1}\text{d}u\int_{r_0}^{r_i}\text{d}r\,\Delta (r)r(r_{\rm i}-r)u^2\left(\frac{\text{d}f}{\text{d}u}\right)^2\left(\frac{\text{d}\Delta}{\text{d}r}\right)^2\nonumber\\
&=-4M\Delta_0\int_{-1}^{+1}\text{d}u\,u^2\left(\frac{\text{d}f}{\text{d}u}\right)^2\int_{r_0/r_{\rm i}}^{1}\text{d}x\,I(x)x(1-x)\left(\frac{\text{d}\ln I(x)}{\text{d}x}\right)^2.
\end{align}
\noindent
The fourth contribution is given by
\begin{align}
\mathcal{E}^{(4)}&=-\int_{-1}^{+1}\text{d}u\int_{r_0}^{r_i}\text{d}r\,\Delta(r)\frac{8M}{r_{\rm i}}(r_{\rm i}-2r)f(T/\Delta)\frac{\partial f(T/\Delta)}{\partial r}\nonumber\\
&=\frac{8M}{r_{\rm i}}\int_{-1}^{+1}\text{d}u\,uf\frac{\text{d}f}{\text{d}u}\int_{r_0}^{r_i}\text{d}r\,\Delta(r)(r_{\rm i}-2r)\frac{\text{d}\ln\Delta}{\text{d}r}\nonumber\\
&=8M\Delta_0\int_{-1}^{+1}\text{d}u\,uf\frac{\text{d}f}{\text{d}u}\int_{r_0/r_{\rm i}}^{1}\text{d}x\,I(x)(1-2x)\frac{\text{d}\ln I(x)}{\text{d}x}.
\end{align}

\noindent
The fifth contribution is:
\begin{align}
\mathcal{E}^{(5)}&=-\int_{-1}^{+1}\text{d}u\int_{r_0}^{r_i}\text{d}r\,\Delta (r)\frac{4M}{r_{\rm i}}r(r_{\rm i}-r)f(T/\Delta)\frac{\partial^2f(T/\Delta)}{\partial r^2}\nonumber\\
&=-\frac{4M}{r_{\rm i}}\int_{-1}^{+1}\text{d}u\,u^2f\frac{\text{d}^2f}{\text{d}u^2}\int_{r_0}^{r_{\rm i}}\text{d}r\,\Delta (r)r(1-r)\left(\frac{\text{d}\ln\Delta}{\text{d}r}\right)^2\nonumber\\
&+\frac{4M}{r_{\rm i}}\int_{-1}^{+1}\text{d}u\,uf\frac{\text{d}f}{\text{d}u}\int_{r_0}^{r_{\rm i}}\text{d}r\,\Delta (r)r(1-r)\left[\frac{\text{d}^2\ln\Delta}{\text{d}r^2}-\left(\frac{\text{d}\ln\Delta}{\text{d}r}\right)^2\right]\nonumber\\
&=-4M\Delta_0\int_{-1}^{+1}\text{d}u\,u^2f\frac{\text{d}^2f}{\text{d}u^2}\int_{r_0/r_{\rm i}}^{1}\text{d}x\,I(x)x(1-x)\left(\frac{\text{d}\ln I(x)}{\text{d}x}\right)^2\nonumber\\
&+4M\Delta_0\int_{-1}^{+1}\text{d}u\,uf\frac{\text{d}f}{\text{d}u}\int_{r_0/r_{\rm i}}^{1}\text{d}x\,I(x)x(1-x)\left[\frac{\text{d}^2\ln I(x)}{\text{d}x^2}-\left(\frac{\text{d}\ln I(x)}{\text{d}x}\right)^2\right].
\end{align}
Taking into account the boundary conditions, this can be also written as
\begin{align}
\mathcal{E}^{(5)}=4M\Delta_0\int_{-1}^{+1}\text{d}u\,u^2\left(\frac{\text{d}f}{\text{d}u}\right)^2\int_{r_0/r_{\rm i}}^1\text{d}x\,I(x)x(1-x)\left(\frac{\text{d}\ln I(x)}{\text{d}x}\right)^2\nonumber\\
+4M\Delta_0\int_{-1}^{+1}\text{d}u\,uf\frac{\text{d}f}{\text{d}u}\int_{r_0/r_{\rm i}}^{1}\text{d}x\,x(1-x)\frac{\text{d}^2I(x)}{\text{d}x^2}.
\end{align}
Then,
\begin{align}
\mathcal{E}^{(3)}+\mathcal{E}^{(4)}+\mathcal{E}^{(5)}&=4M\Delta_0\int_{-1}^{+1}\text{d}u\,uf\frac{\text{d}f}{\text{d}u}\int_{r_0/r_{\rm i}}^{1}\text{d}x\,\left[2(1-2x)\frac{\text{d}I(x)}{\text{d}x}+x(1-x)\frac{\text{d}^2I(x)}{\text{d}x^2}\right]\nonumber\\
&=4M\Delta_0\int_{-1}^{+1}\text{d}u\,uf\frac{\text{d}f}{\text{d}u}\int_{r_0/r_{\rm i}}^{1}\text{d}x\,(1-2x)\frac{\text{d}I(x)}{\text{d}x}+\mathscr{O}(r_0/r_{\rm i}).
\end{align}
The sixth contribution is the following:
\begin{align}
\mathcal{E}^{(6)}&=-\int_{r_0}^{r_i}\text{d}r\,r^2\int_{-\Delta (r)}^{\Delta (r)}\text{d}T\,\frac{2M}{r_{\rm i}r^2}\frac{\displaystyle(3r_{\rm i}-4r)}{\displaystyle\sqrt{2M(1/r-1/r_{\rm i})}}\frac{\partial f(u)}{\partial T}\nonumber\\
&=-\frac{2M}{r_{\rm i}}\int_{-1}^{+1}\text{d}u\,\frac{\text{d}f}{\text{d}u}\int_{r_0}^{r_i}\text{d}r\,\frac{\displaystyle(3r_{\rm i}-4r)}{\displaystyle\sqrt{2M(1/r-1/r_{\rm i})}}\nonumber\\
&=-2\sqrt{\frac{2M}{r_{\rm i}}}\int_{r_0/r_{\rm i}}^{1}\text{d}x\,\frac{\displaystyle(3-4x)\sqrt{x}}{\displaystyle\sqrt{1-x}}\nonumber\\
&=4\sqrt{\frac{2M}{r_{\rm i}}\left(1-\frac{r_0}{r_{\rm i}}\right)\frac{r_0^3}{r_{\rm i}^3}}.
\end{align}
The last contribution to the Einstein-Hilbert action is given by:
\begin{align}
\mathcal{E}^{(7)}&=-\int_{r_0}^{r_i}\text{d}r\,r^2\int_{-\Delta (r)}^{\Delta (r)}\text{d}T\,\frac{4M}{r_{\rm i}r^2}\frac{r(r_{\rm i}-r)}{\displaystyle\sqrt{2M(1/r-1/r_{\rm i})}}\frac{\partial^2f(u)}{\partial T\partial r}\nonumber\\
&=\frac{4M}{r_{\rm i}}\int_{-1}^{+1}\text{d}u\left(\frac{\text{d}f}{\text{d}u}+u\frac{\text{d}^2f}{\text{d}u^2}\right)\int_{r_0}^{r_{\rm i}}\text{d}r\,\frac{r(r_{\rm i}-r)}{\displaystyle\sqrt{2M(1/r-1/r_{\rm i})}}\frac{\text{d}\ln\Delta}{\text{d}r}\nonumber\\
&=0.
\end{align}
In order to write down the Einstein-Hilbert action evaluated on the metric $g_{ab}$ in Eq. \eqref{eq:metricsw2app} in a compact form, we will need the relation
\begin{equation}
\int_{0}^1\text{d}u\,uf\left(\frac{\text{d}f}{\text{d}u}\right)=\frac{1}{2}-\frac{1}{2}\int_{0}^1\text{d}u\,f^2,
\end{equation}
obtained integrating by parts:
\begin{align}
\int_{0}^1\text{d}u\,u\left(\frac{\text{d}f^2}{\text{d}u}\right)=\left.uf^2\right|^1_{0}-\int_{0}^1\text{d}u\,f^2=1-\int_{0}^1\text{d}u\,f^2.
\end{align}
Then,
\begin{align}
\mathcal{E}^{(1)}+\mathcal{E}^{(2)}+\mathcal{E}^{(3)}+\mathcal{E}^{(4)}+\mathcal{E}^{(5)}&=8M\Delta_0\left[\int_{r_0/r_{\rm i}}^1\text{d}x\left(I+\frac{1-2x}{2}\frac{\text{d}I}{\text{d}x}\right)\right]\nonumber\\
&+8M\Delta_0\left[\int_0^1\text{d}u\,f^2(u)\int_{r_0/r_{\rm i}}^1\text{d}x\left(I-\frac{1-2x}{2}\frac{\text{d}I}{\text{d}x}\right)\right]\nonumber\\
&=8M\Delta_0\left[\int_{r_0/r_{\rm i}}^1\text{d}x\left(2I-\frac{1}{2}\right)+\frac{1}{2}\int_0^1\text{d}u\,f^2(u)+\mathscr{O}(r_0/r_{\rm i})\right].
\end{align}
Let us define the constant $k_I=2\int_{r_0/r_{\rm i}}^1\text{d}x\left(2I-1/2\right)$ which, up to $\mathscr{O}(r_0/r_{\rm i})$ and taking into account the monotonous behavior of the function $I(x)$, can only take the values
\begin{equation}
k_I\in\left[1,3\right].
\end{equation}
Putting all together,
\begin{align}
\mathscr{S}_{\rm EH}[g]&=\frac{M\Delta_0}{\sqrt{1-2M/r_{\rm i}}}\left[k_I+\int_0^1\text{d}u\,f^2(u)+\mathscr{O}(r_0/r_{\rm i})\right]\nonumber\\
&+\frac{\sqrt{r_0^3(1-r_0/r_{\rm i})/r_{\rm i}^3}}{\sqrt{2M/r_{\rm i}-1}}.\label{eq:genres1}
\end{align}
This expression is remarkably simple; for instance, it shows no dependence on the derivative of the function $f(u)$ in the interval $u\in[0,1]$. The last piece is independent of $\Delta_0$ and hence irrelevant for the evaluation of the probability amplitude on $\Delta_0$, as it is absorbed in the definition of the normalization factor $\mathscr{N}$ in Eq. \eqref{eq:tamp}.

%------------------------------------
%Bibliography
%------------------------------------
\bibliography{bh-wh_amp-2ndrev}

%------------------------------------
\end{document}